\documentstyle[preprint,aps,epsf]{revtex}
\newcommand{\adots} {{\mathinner{\mkern2mu\raise1pt\hbox{.}\mkern2mu
\raise4pt\hbox{.}\mkern2mu\raise7pt\hbox{.}\mkern1mu}}}

\begin{document}
\title{Dirac monopole with Feynman brackets}

\author{Alain. B\'{e}rard}
\address{LPLI-Institut de Physique, 1 blvd D.Arago, F-57070 Metz, France}

\author{Y. Grandati}
\address{LPLI-Institut de Physique, 1 blvd D.Arago, F-57070 Metz, France}

\author{Herv\'e Mohrbach}
\address{M.I.T, Center for Theoretical Physics, 77 Massachusetts
Avenue,\\ Cambridge, MA 02139-4307 USA\\
and\\
LPLI-Institut de Physique, 1 blvd D.Arago, F-57070 Metz, France}

\maketitle

\begin{abstract}
We introduce the magnetic angular momentum as a consequence of the structure
of the sO(3) Lie algebra defined by the Feynman brackets. The Poincar\'{e}
momentum and Dirac magnetic monopole appears as a direct result of this
framework.
\end{abstract}

\section{Introduction}

In 1990, Dyson\cite{DYSON} published a proof due to Feynman of the Maxwell
equations, assuming only commutation relations between position and
velocity. In this article we don't use the commutation relations explicitly.
In fact what we call a commutation law is a structure of algebra between
position and velocity called in this letter Feynman's brackets. With this
minimal assumption Feynman never supposed the existence of an Hamiltonian or
Lagrangian formalism and didn't need the not gauge invariant momentum.
Tanimura \cite{TANIMURA} extended Feynman's derivation to the case of the
relativistic particle.

In this letter one concentrates only on the following point: the study of a
nonrelativistic particle using Feynman brackets. We show that Poincare's
magnetic angular momentum is the consequence of the structure of the sO(3)
Lie algebra defined by Feynman's brackets.

\section{Feynman brackets}

Assume a particle of mass m moving in a three dimensional Euclidean space
with position: $x_{i}(t)$ ($i=1,2,3$) depending on time. As Feynman we
consider a non associative internal structure (Feynman brackets) between the
position and the velocity. The starting point is the bracket between the
various components of the coordinate:

\begin{equation}
\lbrack x_{i},x_{j}]=0  \label{un}
\end{equation}
We suppose that the brackets have the same properties than in Tanimura's
article \cite{TANIMURA}, that is:

\begin{equation}
\lbrack A,B]=-[A,B]
\end{equation}

\begin{equation}
\lbrack A,BC]=[A,B]C+[A,C]B
\end{equation}

\begin{equation}
\frac{d}{dt}[A,B]=[\stackrel{.}{A},B]+[A,\stackrel{.}{B}]
\end{equation}
where the arguments $A$, $B$ and $C$ are the positions or the
velocities\medskip .

The following Jacobi identity between positions is also trivially satisfied:

\begin{equation}
\lbrack x_{i},[x_{j},x_{k}]]+[x_{j},[x_{k},x_{i}]]+[x_{k},[x_{i},x_{j}]]=0
\end{equation}
In addition we will need also a ``Jacobi identity'' mixing position and
velocity such that: 
\begin{equation}
\lbrack \stackrel{.}{x_{i}},[\stackrel{.}{x_{j}},x_{k}]]+[\stackrel{.}{x_{j}}%
,[x_{k},\stackrel{.}{x_{i}}]]+[x_{k},[\stackrel{.}{x_{i}},\stackrel{.}{x_{j}}%
]]=0  \label{deux}
\end{equation}
Deriving (\ref{un}) gives:

\begin{equation}
\lbrack \stackrel{.}{x_{i}},x_{j}]+[x_{i},\stackrel{.}{x_{j}}]=0
\end{equation}
This implies:

\begin{equation}
\lbrack x_{i},\stackrel{.}{x_{j}}]=g_{ij}(x_{k}),  \label{trois}
\end{equation}
where $g_{ij}(x_{k})$ is a symmetric tensor. We consider here only the case
where: 
\begin{equation}
g_{ij}=\frac{\delta _{ij}}{m}
\end{equation}
this gives the following relations: 
\begin{equation}
\lbrack x_{i},f(x_{j})]=0
\end{equation}
\begin{equation}
\lbrack x_{i},f(x_{j},\stackrel{.}{x_{j}})]=\frac{1}{m}\frac{\partial f(%
\stackrel{.}{x_{j}})}{\partial \stackrel{.}{x_{i}}}
\end{equation}
\begin{equation}
\lbrack \stackrel{.}{x}_{i},f(x_{j})]=-\frac{1}{m}\frac{\partial f(x_{j})}{%
\partial x_{i}}
\end{equation}

\medskip

\section{Angular momentum}

Suppose first the following relation:

\begin{equation}
\lbrack \stackrel{.}{x}_{i},\stackrel{.}{x_{j}}]=0
\end{equation}
which permits to say that the force law is velocity independent: 
\begin{equation}
\stackrel{..}{x}_{i}=\stackrel{..}{x}_{i}(x_{j})
\end{equation}
By definition the orbital angular momentum is:

\begin{equation}
L_{i}=m\varepsilon _{ijk}x_{j}\stackrel{.}{x_{k}}
\end{equation}
which satisfies the standard sO(3) Lie algebra for Feynman's brackets:

\begin{equation}
\lbrack L_{i,}L_{j}]=\varepsilon _{ijk}L_{k}  \label{quatre}
\end{equation}
The transformation law of the position and velocity under this symmetry is:

\begin{equation}
\lbrack x_{i,}L_{j}]=\varepsilon _{ijk}x_{k}  \label{cinq}
\end{equation}

\begin{equation}
\lbrack \stackrel{.}{x_{i}},L_{j}]=\varepsilon _{ijk}\stackrel{.}{x_{k}}
\label{six}
\end{equation}
We consider as Feynman \cite{DYSON}, the case with a ''gauge curvature'':

\begin{equation}
\lbrack \stackrel{.}{x}_{i},\stackrel{.}{x_{j}}]=\frac{\alpha }{m^{2}}F_{ij}
\label{feyn}
\end{equation}
where $F$ must be an antisymmetric tensor (electromagnetic tensor for our
example) and $\alpha $ a constant. The goal of our work is to see what
happens if we keep the structure of the Lie algebra of the angular momentum
and the transformation law of the position and velocity. Using (\ref{deux})
we get the relations:

\begin{eqnarray}
\alpha \frac{\partial F_{jk}}{\partial \stackrel{.}{x}_{i}} &=&-m^{2}[x_{i},[%
\stackrel{.}{x_{j}}\stackrel{.}{,x_{k}}]] \\
&=&-m^{2}[\stackrel{.}{x_{j}},[x_{_{i},}\stackrel{.}{x}_{k}]]+[\stackrel{.}{x%
}_{k},[\stackrel{.}{x}_{j,}x_{i}]]=0  \nonumber
\end{eqnarray}
then the electromagnetic tensor is independent of the velocity:

\begin{equation}
F_{jk}=F_{jk}(x_{i})
\end{equation}
\textit{\ }By deriving (\ref{trois}) we have:

\begin{equation}
\lbrack x_{i},\stackrel{..}{x}_{j}]=-[\stackrel{.}{x}_{i},\stackrel{.}{x_{j}}%
]=-\frac{\alpha F_{ij}}{m^{2}}
\end{equation}
then:

\begin{equation}
m\frac{\partial \stackrel{..}{x}_{j}}{\partial \stackrel{.}{x_{i}}}=\alpha
F_{ji}(x_{k})
\end{equation}
or:

\begin{equation}
m\stackrel{..}{x}_{i}=\alpha (E_{i}(x_{k})+F_{ij}(x_{k})\stackrel{.}{x}_{j})
\end{equation}
We get the '' Lorentz force's law'', where the electric field appears as a
constant of integration (this is not the case for the relativistic problem,
see \cite{TANIMURA}). Now the force law is velocity dependent: 
\begin{equation}
\stackrel{..}{x}_{i}=\stackrel{..}{x}_{i}(x_{j},\stackrel{.}{x}_{j})
\end{equation}

For the case (\ref{feyn}), the equations (\ref{quatre}), (\ref{cinq})and (%
\ref{six}) become :

\begin{equation}
\lbrack x_{i,}L_{j}]=\varepsilon _{ijk}x_{k}
\end{equation}

\begin{equation}
\lbrack \stackrel{.}{x}_{i,}L_{j}]=\varepsilon _{ijk}\stackrel{.}{x}%
_{k}+\alpha \varepsilon _{jkl}x_{k}\frac{F_{il}}{m}
\end{equation}

\begin{equation}
\lbrack L_{i,}L_{j}]=\varepsilon _{ijk}L_{k}+\alpha \varepsilon
_{ikl}\varepsilon _{jms}x_{k}x_{m}F_{ls}
\end{equation}
Introducing the magnetic field we write $F$ in the following form:

\begin{equation}
F_{ij}=\varepsilon _{ijk}B_{k},
\end{equation}
We get then the new relations:

\begin{equation}
\lbrack \stackrel{.}{x}_{i,}L_{j}]=\varepsilon _{ijk}\stackrel{.}{x}_{k}+%
\frac{\alpha }{m}\{x_{i}B-\delta _{ij}(\stackrel{\rightarrow }{r}.\stackrel{%
\rightarrow }{B})\}
\end{equation}

\begin{equation}
\lbrack L_{i,}L_{j}]=\varepsilon _{ijk}\{L_{k}+\alpha x_{k}(\stackrel{%
\rightarrow }{r}.\stackrel{\rightarrow }{B})\}
\end{equation}
To \medskip keep the standard relations we introduce a generalized angular
momentum:

\begin{equation}
\mathcal{L}_{i}=L_{i}+M_{i}
\end{equation}
We call $M_{i}$ the magnetic angular momentum because it depends on the
field $\stackrel{\rightarrow }{B}$. It has no connection with the spin of
the particle, which can be introduced by looking at the spinorial
representations of the sO(3) algebra. Now we impose for the $\left\{ \alpha
_{j}\right\} $'s the following commutation relations: 
\begin{equation}
\lbrack \stackrel{.}{x}_{i,}\mathcal{L}_{j}]=\varepsilon _{ijk}x_{k}
\end{equation}
\begin{equation}
\lbrack \stackrel{.}{x}_{i,}\mathcal{L}_{j}]=\varepsilon _{ijk}\stackrel{.}{x%
}_{k}  \label{lx}
\end{equation}
\begin{equation}
\lbrack \mathcal{L}_{i,}\mathcal{L}_{j}]=\varepsilon _{ijk}\mathcal{L}_{k}
\label{l1}
\end{equation}
This first relation gives:

\begin{equation}
M_{i}=M_{i}(x_{j})
\end{equation}
and the second:

\begin{equation}
\lbrack \stackrel{.}{x}_{i,}M_{j}]=\frac{\alpha }{m}[\delta _{ij}(\stackrel{%
\rightarrow }{r}.\stackrel{\rightarrow }{B})-x_{i}B_{j}]
\end{equation}
If we replace it in (\ref{l1}) we deduce:

\begin{equation}
M_{i}=-\alpha (\stackrel{\rightarrow }{r}.\stackrel{\rightarrow }{B})x_{i}
\end{equation}
Putting this result in (\ref{lx}) gives the following equation of constraint
for the field $\overrightarrow{B}:$

\begin{equation}
x_{i}B_{j}+x_{j}B_{i}=-x_{j}x_{k}\frac{\partial B_{k}}{\partial x_{i}}
\end{equation}
One solution has the form of a radial vector field centered at the origin:

\begin{equation}
\stackrel{\rightarrow }{B}=\beta \frac{\stackrel{\rightarrow }{r}}{r^{3}}
\end{equation}
The generalized angular momentum then becomes:

\begin{equation}
\stackrel{\rightarrow }{\mathcal{L}}=m(\stackrel{\rightarrow }{r}\wedge 
\stackrel{\stackrel{.}{\rightarrow }}{r})-\alpha (\stackrel{\rightarrow }{r}.%
\stackrel{\rightarrow }{B})\stackrel{\rightarrow }{r}
\end{equation}
We can check the conservation of the total angular momentum:

\begin{equation}
\frac{d\stackrel{\rightarrow }{\mathcal{L}}}{dt}=m(\stackrel{\rightarrow }{r}%
\wedge \stackrel{\stackrel{..}{\rightarrow }}{r})-\alpha \{\stackrel{%
\rightarrow }{r}\wedge (\stackrel{.}{\overrightarrow{r}}\wedge 
\overrightarrow{B})\}=0
\end{equation}
because the particle satisfies the usual equation of motion:

\begin{equation}
m\frac{d^{2}\stackrel{\stackrel{..}{\rightarrow }}{r}}{dt^{2}}=\alpha (%
\stackrel{\stackrel{.}{\rightarrow }}{r}\wedge \stackrel{\rightarrow }{B})
\end{equation}
If we choose: $\alpha =q$ and $\beta =g$, where $q$ and $g$ are the electric
and magnetic charges, we obtain as a the special case the Poincar\'{e} \cite
{POINCARE} magnetic angular momentum:

\begin{equation}
\stackrel{\rightarrow }{M}=-\frac{qg}{4\pi }\frac{\stackrel{\rightarrow }{r}%
}{r}
\end{equation}
and the Dirac \cite{DIRAC} magnetic monopole:

\begin{equation}
\stackrel{\rightarrow }{B}=\frac{g}{4\pi }\frac{\stackrel{\rightarrow }{r}}{%
r^{3}}
\end{equation}
In addition we find that for the Dirac monopole the source of the field is
localized at the origin:

\begin{equation}
div\overrightarrow{B}=[\stackrel{.}{x_{i}},[\stackrel{.}{x_{j}},\stackrel{.}{%
x}_{k}]]+[\stackrel{.}{x_{j}},[\stackrel{.}{x}_{k},\stackrel{.}{x_{i}}]]+[%
\stackrel{.}{x}_{k},[\stackrel{.}{x_{i}},\stackrel{.}{x_{j}}]]=\frac{g}{4\pi 
}[\stackrel{.}{x}_{i},\frac{x_{i}}{r^{3}}]=g\delta (\stackrel{\rightarrow }{r%
})
\end{equation}
We see that in the construction of the Feynman's brackets algebra the fact
that we didn't impose the Jacobi identity between the velocities is a
necessary condition to obtain a monopole solution.

\medskip

In summary, we used the Feynman's algebra between position and velocity to
compute the algebra of the angular momentum of a non relativistic particle
in a electromagnetic field. The Dirac monopole and magnetic angular momentum
is a direct consequence of the conservation of the form of the standard
sO(3) Lie algebra.

\medskip

\medskip

\section{Casimir Operator}

In the same spirit, it is interesting to introduce $L^{2},$ the Casimir
operator of sO(3) Lie algebra. Again we want to keep the same commutation
relations in the two cases corresponding to zero and non zero curvature.

In the first case, we easily see that:

\begin{equation}
\lbrack x_{i,}L^{2}]=2(\stackrel{\rightarrow }{L}\wedge \stackrel{%
\rightarrow }{r})_{i}
\end{equation}

\begin{equation}
\lbrack \stackrel{.}{x}_{i,}L^{2}]=2(\stackrel{\rightarrow }{L}\wedge 
\stackrel{\stackrel{.}{\rightarrow }}{r})_{i}
\end{equation}

\begin{equation}
\lbrack L_{i,}L^{2}]=0
\end{equation}
and in presence of a curvature:

\begin{equation}
\lbrack x_{i,}L^{2}]=2(\stackrel{\rightarrow }{L}\wedge \stackrel{%
\rightarrow }{r})_{i}
\end{equation}

\begin{equation}
\lbrack \stackrel{.}{x}_{i,}L^{2}]=2[(\stackrel{\rightarrow }{L}\wedge 
\stackrel{\stackrel{.}{\rightarrow }}{r})_{i}+\alpha (\stackrel{\rightarrow 
}{L}\wedge \stackrel{\rightarrow }{r})_{l}F_{il}]
\end{equation}

\begin{equation}
\lbrack L_{i,}L^{2}]=2\alpha (\stackrel{\rightarrow }{L}\wedge \stackrel{%
\rightarrow }{r})_{i}(\stackrel{\rightarrow }{r}.\stackrel{\rightarrow }{B})
\end{equation}
then we want: 
\begin{equation}
\lbrack x_{i,}\mathcal{L}^{2}]=2(\stackrel{\rightarrow }{\mathcal{L}}\wedge 
\stackrel{\rightarrow }{r})_{i}
\end{equation}
\begin{equation}
\lbrack \stackrel{.}{x}_{i,}\mathcal{L}^{2}]=2(\stackrel{\rightarrow }{%
\mathcal{L}}\wedge \stackrel{\stackrel{.}{\rightarrow }}{r})_{i}
\end{equation}
\begin{equation}
\lbrack \mathcal{L}_{i,}\mathcal{L}^{2}]=0
\end{equation}
and we can deduce:

\begin{equation}
\lbrack x_{i,}M^{2}]=2(\stackrel{\rightarrow }{M}\wedge \stackrel{%
\rightarrow }{r})_{i}
\end{equation}
\begin{equation}
\lbrack \stackrel{.}{x}_{i,}M^{2}]=2[(\stackrel{\rightarrow }{M}\wedge 
\stackrel{\rightarrow }{r})_{i}-\alpha (\stackrel{\rightarrow }{L}\wedge 
\stackrel{\rightarrow }{r})_{l}F_{il}
\end{equation}

\begin{equation}
2\alpha (\stackrel{\rightarrow }{L}\wedge \stackrel{\rightarrow }{r})_{i}(%
\stackrel{\rightarrow }{r}.\stackrel{\rightarrow }{B}%
)+[L_{i},M^{2}]+[M_{i,}L^{2}]=0
\end{equation}
The last equation becomes after a straightforward computation:

\medskip 
\begin{equation}
(\stackrel{\rightarrow }{M}\wedge \stackrel{\rightarrow }{r})(\stackrel{%
\rightarrow }{L}\wedge \stackrel{\stackrel{.}{\rightarrow }}{r})-(\stackrel{%
\rightarrow }{L}\wedge \stackrel{\rightarrow }{r})(\stackrel{\rightarrow }{M}%
\wedge \stackrel{\stackrel{.}{\rightarrow }}{r})-(\stackrel{\rightarrow }{M}%
\wedge \stackrel{\stackrel{.}{\rightarrow }}{r})(\stackrel{\rightarrow }{L}%
\wedge \stackrel{\rightarrow }{r})+(\stackrel{\rightarrow }{M}\wedge 
\stackrel{\rightarrow }{r})(\stackrel{\rightarrow }{L}\wedge \stackrel{%
\stackrel{.}{\rightarrow }}{r})=0
\end{equation}
We can check that this equation of constraint is in particular satisfied for
the Poincar\'{e} angular momentum.

\section{Conclusion}

We find that the structure of Feynman's brackets (without an Hamiltonian or
Lagrangian), illuminates the connections between the spaces with gauge
curvature, the sO(3) Lie algebra and the existence of the Poincar\'{e}
magnetic angular momentum. It seems that more than the phase space
formalism, the Feynman's one is a good approach of the mechanics in a space
with gauge symmetry, because it avoids the introduction of the not gauge
invariant momentum. Further, other applications of this method, for example,
the case of the Minkowski space with Lorentz Lie algebra, will be consider
in the future.

\end{document}